# A note on radial basis function computing

## W. Chen[*] and W. He[**]

[*]*Department of Mechanical System Engineering, Shinshu University, Wakasato 4-17-1, Nagano City, Nagano 380-8533, Japan (chenw@homer.shinshu-u.ac.jp)*

[**]*Department of Electrical Engineering, Jiangsu University of Science and Technology, Zhenjiang City, Jiangsu Province 212013, P. R. China.*

## 1. Introduction

This note carries three purposes involving our latest advances on the radial basis function (RBF) approach. First, we will introduce a new scheme employing the boundary knot method (BKM) [1] to nonlinear convection-diffusion problem. It is stressed that the new scheme directly results in a linear BKM formulation of nonlinear problems by using response point-dependent RBFs, which can be solved by any linear solver. Then we only need to solve a single nonlinear algebraic equation for desirable unknown at some inner node of interest. The numerical results demonstrate high accuracy and efficiency of this nonlinear BKM strategy. Second, we extend the concepts of distance function, which include time-space and variable transformation distance functions. Finally, we demonstrate that if the nodes are symmetrically placed, the RBF coefficient matrices have either centrosymmetric or skew centrosymmetric structures. The factorization features of such matrices lead to a considerable reduction in the RBF computing effort. A simple approach is also presented to reduce the ill-conditioning of RBF interpolation matrices in general cases.

## 2. BKM linear formulation of nonlinear problems

Chen and Tanaka [1] found that if only boundary knots are used, the BKM can formulate linear analogization of nonlinear differential equations with linear boundary conditions. Consider the Burger-like convection-diffusion equation [2]

$$\nabla^2 u - u_x u = 0 \qquad (1)$$

with inhomogeneous boundary condition

$$u = -2 / x \qquad (2)$$

Eq. (2) is also a particular solution of Eq. (1). The geometry of the problem is an ellipse with semi-major axis of length 2 and semi-minor axis of length 1. Note that the origin of the Cartesian co-ordinates system is dislocated to the node (3,0) to circumvent singularity at $x=0$. Using the scheme given in [1], Eq. (1) is restated as

$$\nabla^2 u + u = u + u_x u. \qquad (3)$$

When only the boundary nodes are employed, the resulting BKM formulation will be a linear algebraic equation in terms of 2D Helmhotz non-singular general solution. For details of such BKM procedure see [1]. The numerical solutions of this normal BKM procedure have the average relative errors 3.9e-2 for $N=5$, 1.1e-1 for $N=9$, 1.4e-1 for $N=13$ at some inner nodes, where $N$ is the number of boundary nodes used. It is noted that the performances are unstable and solutions inaccurate. If we use interior points in the BKM, the accuracy and stability will be improved greatly at the expense of sacrificing linear formulation as in the dual reciprocity BEM (DRM) [2]. It is note that the convection term rather than nonlinear constitution here causes the deficiency of the BKM solutions if not using inner points.

In the DRM, it is reported that the use of the fundamental solution of convection diffusion equation can significantly improve the solution accuracy of transient convection diffusion problem with only one inner point. This suggests us that a non-singular general solution of convection-diffusion operator may be much more suitable for this problem. By analogy with the fundamental solution of the convection-diffusion equation, the response knot-dependent non-singular general solution of Eq. (1) is given by

$$u(r, x) = e^{-u(x-x_s)/2} \phi\left(\frac{|u|}{\sqrt{2}} r\right), \qquad (4)$$

where radial function $\varphi$ is the zero order Bessel $J_o$ or modified Bessel function $I_o$ of the first kind dependent on the sign of flow velocity $u$. In this case it is the latter. The RBF approximation is given by

$$u_i = \sum_{k=1}^{N} \alpha_k e^{-u_i(x_i-x_s)/2} I_0\left(\frac{|u_i|}{\sqrt{2}} r_{ik}\right). \qquad (5)$$

Note that the above RBF representation differs from the normal one in that we here use the response point-dependent RBFs even if we do not know the value of $u$. In terms of the BKM using only boundary knots, Eq. (1) is analogized by substituting boundary conditions into Eq. (5), namely,

$$\sum_{k=1}^{N} \alpha_k e^{(x_i-x_s)/x_i} I_0\left(r_{ik}\sqrt{2}/x_i\right) = -2/x_i. \qquad (6)$$

The formulation (6) is a set of simultaneous linear algebraic equations and can be solved easily by any linear solver.

Then we can evaluate the value of $u$ at any inner nodes of interest through the solution of a single nonlinear equation (5). Note that the RBF expansion coefficients in Eq. (5) are now already evaluated from Eq. (6) and the only one unknown is the value of $u$ at a specified single inner knot. This study used simple bisection method to handle such a single nonlinear equation. There are no concerns here relating to the expensive repeated evaluation and inverse of Jacobian matrix, stability issue and the careful guess of initial solutions.

In summary, the present nonlinear BKM scheme can be viewed as a two-step procedure. First, the linear BKM formulation of nonlinear problems is produced using response node-dependent RBFs. Then, the second step is to calculate the solution at any inner node of interest through solving a single nonlinear algebraic equation.

Table 1. Relative errors for BKM linear formulation of Burger-like equation

| x | y | DRM(33) | BKM (9) | BKM (11) |
|---|---|---|---|---|
| 4.5 | 0.0 | 2.3e-3 | 2.8e-3 | 2.5e-3 |
| 4.2 | -0.35 | 2.1e-3 | 2.3e-3 | 2.9e-3 |
| 3.6 | -0.45 | 5.4e-3 | 4.4e-3 | 6.2e-3 |
| 3.0 | -0.45 | 4.5e-3 | 1.0e-2 | 9.2e-3 |
| 2.4 | -0.45 | 1.2e-3 | 1.2e-2 | 5.7e-3 |
| 1.8 | -0.35 | 9.0e-4 | 7.0e-3 | 3.2e-3 |
| 3.9 | 0.0 | 3.9e-3 | 4.1e-3 | 5.5e-3 |
| 3.3 | 0.0 | 3.3e-3 | 9.1e-3 | 1.0e-2 |
| 3.0 | 0.0 | 4.5e-3 | 1.2e-2 | 1.1e-2 |
| 2.7 | 0.0 | 2.7e-3 | 1.4e-2 | 1.1e-2 |
| 2.1 | 0.0 | 3.2e-3 | 1.1e-2 | 4.9e-3 |

Table 1 lists the BKM results compared with the solutions of the dual reciprocity BEM [2]. Average relative errors of the present BKM for $N$=9, 11 are respectively 8.5e-3, 7.5e-3. Compared with the previous BKM procedure, the solution accuracy is significantly improved while still keeping the linear BKM formulation of nonlinear problems. However, it is noted that the solution accuracy is still not always improved with incremental number of boundary knots. For example, average relative errors for $N$=13, 15, 17, 19, 21 are respectively 8.3e-3, 8.3e-3, 8.8e-3, 8.9e-3, 1.9e-2. Namely, the highest average accuracy is achieved for N=11. As in the other global collocation techniques, this is due to the ill-conditioning system matrix for large number of knots. Anyway overall solution procedure is rather stable. The accuracy and efficiency of the present BKM scheme are very encouraging.

On the other hand, the solutions of the DRM [2] were obtained with 16 boundary nodes and 17 inner points. Therefore, it is not surprising that the DRM solutions [2] are slightly more accurate than the present BKM ones. It is also noted that the DRM formulation is a set of simultaneous nonlinear algebraic equations. The programming, computing effort and storage requirements in the DRM are much higher than the BKM. The present case only involves the Dirichlet conditions. For more complicated boundary conditions, the resulting BKM formulation may not be a set of linear algebraic equations even if we only employ boundary knots. By any measure, however, the size of the BKM analogous equations will be much smaller than that of the DRM.

The essential idea behind this work may be extended to the BEM, DRM, method of fundamental solution, and multiple reciprocity BEM for nonlinear problems. Namely, as in handling linear varying parameter problems, the response knot-dependent fundamental solution may be employed in the nonlinear computing even if some parameters of fundamental solution are unknown.

## 3. Redefinition of distance functions

### 3.1. Time-space distance

Golberg and Chen [3] summarized the Euclidean and geodisc distance function as the two kinds of distance functions used in the RBF. Chen and Tanaka [1] introduced a time-space distance function to eliminate time dependence within the framework of the RBF numerical schemes. This study further develops this work in combination with the Green RBF [1] to create efficient operator-dependent RBFs. Note that the Green RBF was named after the general solution RBF (GS-RBF) in [1]. Here we feel it is more proper to call it as Green RBF since it uses both the fundamental solution of the related operator and inhomogeneous terms based on the known Green second identity.

First, let us consider the equation governing wave propagation

$$\nabla^2 u = \frac{1}{c^2} u_{tt} + f(p, t). \qquad (7)$$

Let

$$s = ict, \qquad (8)$$

where $p$ denotes multidimensional variable and $i$ means unit imaginary number, we have

$$\nabla^2 u + u_{ss} = f(p, t). \qquad (9)$$

By analogy with the Euclidean definition of distance variable, the time-space distance function is defined

$$r_j = \sqrt{r_{pj}^2 + \left(s - s_j\right)^2} = \sqrt{r_{pj}^2 - c^2 \left(t - t_j\right)^2}, \quad (10)$$

where $r_{pj}$ denotes the normal spatial Euclidean distance function. The above definition can lead to complex distance variable due to the presence of minus operation. It is safe to use

$$r_j = \sqrt{r_{pj}^2 + c^2 \Delta t_j^2}. \qquad (11a)$$

$$r_j = \sqrt{c^2 \Delta t_j^2 - r_{pj}^2} H\left(c \Delta t_j - r_{pj}^2\right). \qquad (11b)$$

$$r_j = r_{pj} - c \Delta t_j \ \ \& \ \ r_j = r_{pj} + c \Delta t_j. \qquad (11c)$$

The definition (11a,b,c) of distance function differ from the standard radial distance function in that the time variable is handled equally as the space variables. Note that distances (11a) are respectively used in two RBFs for one problem. In general, hyperbolic and elliptic equations have solutions whose arguments have the form $p+at$ and $p+ibt$ respectively, where $a$ and $b$ are real. Namely,

$$u(p, t) = f(\zeta) \qquad (12)$$

where $\zeta$ is some linear combination of $p$ and $t$. This provides some theoretical support to use time-space distance functions (10) and (11). However, the above situations do not hold for

parabolic-type diffusion equation

$$\nabla^2 u = \frac{1}{k} u_t + f(p,t). \qquad (13)$$

Its transient fundamental solution is well known

$$u^* = \frac{1}{(t_j - t)^{d/2}} \exp\left(-\frac{r_p^2}{4k(t_j - t)}\right) H(t_j - t), \quad (14)$$

where $d$ is the space dimensionality, $H$ is the Heaviside function. We can define the corresponding time-space RBF (TS-RBF) as

$$\phi(r_p, t, t_j) = h(r_p, t, t_j) u^*(r, t, t_j), \qquad (15)$$

Here $h$ is chosen according to problem feature. It is stressed that in this case the response and source nodes must be totally staggered to avoid singularity in time dimension.

On the other hand, Chen et al. [4] proposed time-space non-singular general solution

$$u^* = A e^{-k(t-t_j)} \phi(r_p), \qquad (16)$$

for diffusion problem and

$$u^* = \left[C \cos\left(c(t - t_j)\right) + D \sin\left(c(t - t_j)\right)\right] \phi(r_p) \quad (17)$$

for wave problems, where $\varphi(r_p)$ is the zero order Bessel function of the first kind for 2D problems and $\sin(r_p)/r_p$ for 3D problems. For example, consider the free symmetrical vibration of a very large membrane governed by the equation

$$\frac{\partial^2 z}{\partial r^2} + \frac{1}{r}\frac{\partial z}{\partial r} = \frac{1}{c^2}\frac{\partial^2 z}{\partial t^2} \qquad (18)$$

with $z=f(r)$, $\partial z/\partial t = g(r)$ when $t=0$. We have the solution [5]

$$z(r,t) = \int_0^\infty \xi \bar{f}(\xi)\cos(\xi ct) J_0(\xi r) d\xi + \frac{1}{c}\int_0^\infty \bar{g}(\xi)\sin(\xi ct) J_0(\xi r) d\xi \qquad (19)$$

where upper-dashed $f$ and $g$ are the zero-order

Hankel transforms of $f(r)$ and $g(r)$, respectively. In terms of the Green RBF [1], we have

$$z(r,t) = \left[A\cos\left(c(t - t_j)\right) + B\sin\left(c(t - t_j)\right)\right] J_0(r). \quad (20)$$

Substituting the non-singular general solutions (16), (17) and (20) into Eq. (15) produces the TS-RBF without singularity.

The time-space distance function and corresponding TS-RBF are also expected to be applicable to transient data processing such as motion picture and movie.

### 3.2 Varying parameter problems

The Green-RBF is constructed based on the canonical form of some operators such as the known Laplace or Helmholtz operators. However, many engineering problems do not possess such standard form operators. This section will show that some cares may be taken to handle these problems.

The general second order partial differential system with varying coefficient can be stated as

$$R\frac{\partial^2 u}{\partial x^2} + S\frac{\partial^2 u}{\partial x \partial y} + T\frac{\partial^2 u}{\partial y^2} = 0, \qquad (21)$$

where $R$, $S$ and $T$ are continuous functions of $x$ and $y$. We can translate it into the canonical Laplacian by a suitable change of independent variables [5]

$$\xi = f_1(x, y), \quad \eta = f_2(x, y). \qquad (22)$$

The corresponding distance function of the Euclidean norm is given by

$$r = \sqrt{(\xi - \xi_j)^2 + (\eta - \eta_j)^2}. \qquad (23)$$

Substituting Eq. (22) into Eq. (23), we define the distance function in terms of the original independent variables of $x$ and $y$ as

$$r = \sqrt{\left[f_1(x, y) - f_1(x_j, y_j)\right]^2 + \left[f_2(x, y) - f_2(x_j, y_j)\right]^2}. \qquad (24)$$

In the following we illustrate one special case. Let us consider [6]

$$y^m \frac{\partial^2 u}{\partial x^2} + \frac{\partial^2 u}{\partial y^2} = 0, \quad (y \geq 0, m \succ -2), \qquad (25)$$

its general solution is

$$u = r_2^{-2\beta} w(r^2), \qquad (26)$$

where $\beta = m/2(m+2)$, $w$ is the hypergeometric functions,

$$r = \frac{r_1}{r_2}, \qquad (27a)$$

$$r_1 = \sqrt{\left(x - x_j\right)^2 + \frac{4}{(m+2)^2}\left(y^{\frac{m+2}{2}} - y_j^{\frac{m+2}{2}}\right)^2}, \quad (27b)$$

$$r_2 = \sqrt{\left(x - x_j\right)^2 + \frac{4}{(m+2)^2}\left(y^{\frac{m+2}{2}} + y_j^{\frac{m+2}{2}}\right)^2}. \quad (27c)$$

Let

$$\xi = x, \quad \eta = \frac{2}{m+2} y^{\frac{m+2}{2}}, \qquad (28)$$

we have

$$\frac{\partial^2 u}{\partial \xi^2} + \frac{\partial^2 u}{\partial \eta^2} = 0. \qquad (29)$$

Substituting Eq. (28) into Eq. (23) also yields the distance function (27b). Similar situation holds for another example

$$\frac{\partial^2 u}{\partial x^2} + y \frac{\partial^2 u}{\partial y^2} + \alpha \frac{\partial u}{\partial y} = 0, \quad (y \geq 0), \qquad (30)$$

where $\alpha$ is a coefficient.

The above analyses show that for one certain problem, we can use multiple different definitions of distance variable simultaneously such as Eqs. (27a,b,c). The use of RBF should fully consider the features of the targeted problems.

### 3.3. Wavelet RBF

Chen and Tanaka [1] constructed pre-wavelet RBFs with $\sqrt{r_j^2 + c_j^2}$ instead of $r$ into the RBFs, where $c_j$ is dilution parameters. For example, numerical experiments with pre-wavelet TPS $r_j^{2m} \ln\sqrt{r_j^2 + c_j^2}$ manifests spectral convergence as in the multiquadratic (MQ). This work can be generalized by

$$u = \sum_{k=1}^{N} c_k \phi\left(\lambda_k r_k + d_k\right), \qquad (31)$$

where $\lambda$ and $d$ are respectively dilate and location coefficients of the wavelet. . is the RBF which can be here seen as a wavelet parent function. Such wavelet-like RBF may be especially attractive for adaptable handling geometry singularity and localized shock-like solutions due to its inherent multiscale feature combined with a spatial localization.

Fasshauer and Schumaker [7] summarized some wavelets using sphere RBFs. It will be beneficial to pay more attentions on this aspect, especially for creating orthonormal wavelet-like RBF.

## 4. Coefficient matrix structures and ill-conditioning

It is well known that the RBF interpolation matrix has symmetric structure irrespective of the geometry and node placements. This study will show that the RBF matrix also carries the centrosymmetric structure if the nodes are symmetrically placed. Let us consider a $d$-dimension space problem. The Euclidean distance norm is defined as

$$r_{ij} = \sqrt{\sum_{k=1}^{d} \left(x_i^{(k)} - x_j^{(k)}\right)^2}. \qquad (32)$$

The symmetric node spacing is understood as

$$x_i^{(k)} + x_{N+1-i}^{(k)} = x_j^{(k)} + x_{N+1-j}^{(k)} = c^{(k)}, \qquad (33)$$

where $c^{(k)}$ are constants. Thus, it is obvious

$$r_{ij} = r_{N+1-i, N+1-j}. \qquad (34)$$

Based on Eq. (34), one can easily verify that the

RBF coefficient matrices have centrosymmetric structure as shown in Eq. (34) for even order derivative and skew centrosymmetric structure as shown below

$$a_{ij} = -a_{N+1-i,N+1-j} \qquad (35)$$

for odd order derivative, where $a$ denotes an entry of a matrix. Therefore, if the nodes are symmetrically placed, the RBF coefficient matrices for derivatives have either symmetric centrosymmetric or skew symmetric centrosymmetric structures.

Centrosymmetric matrices can easily be decomposed into two half-sized matrices. Such factorization merit leads to a considerable reduction in computing effort for determinant, inversion and eigenvalues. For more related details see [8-9]. On the other hand, even if the total node spacing is not symmetric, such decomposition processing can still effectively reduce the ill-conditioning of the resulting RBF matrices by preconditioning

$$\hat{A} = \begin{bmatrix} I & -J \\ I & J \end{bmatrix} A \begin{bmatrix} I & I \\ -J & J \end{bmatrix} \qquad (36)$$

where $A$ means a RBF matrix of even order, $J$ is contra-identity matrix. A similar but distinct preconditioning transform matrix exists for odd order matrix.

In addition, we observe that as in the traditional collocation method [10], very large entries of RBF derivative matrices appear in the upper and lower two rows and middle columns and largely account for the ill-conditioning of large-size RBF system. Accordingly some elementary matrix transformations can simply significantly reduce the RBF ill-conditioning with very little effort.